\documentclass[sigconf,nonacm]{acmart}

\usepackage{booktabs}
\usepackage{listings}
\usepackage{xcolor}

\lstdefinelanguage{sexp}{
  morekeywords={intent,design-constraints,pillar,component,symbols,function,sig,purpose,invariants,data-flow,role,communication,db-access,must-use,only},
  morestring=[b]",
  morecomment=[l]{;;},
  sensitive=true,
}
\renewcommand\footnotetextcopyrightpermission[1]{}
\begin{document}

\title{Formal Architecture Descriptors as Navigation Primitives\\for AI Coding Agents}

\author{Ruoqi Jin}
\email{jinruoqi@xiaojinpro.com}
\affiliation{%
  \institution{Independent Researcher}
  \country{China}
}

\begin{abstract}
AI coding agents spend a substantial fraction of their tool calls on undirected codebase exploration. We investigate whether providing agents with formal architecture descriptors---structured documents declaring module boundaries, symbol signatures, constraints, and data flows---can reduce this navigational overhead.

We present three complementary studies. First, a controlled experiment (24 code localization tasks $\times$ 4 conditions, Claude Sonnet 4.6, temperature=0) demonstrates that architecture context---regardless of format---reduces navigation steps by 33--44\% (Wilcoxon signed-rank $p = 0.009$, Cohen's $d = 0.92$). No significant format difference was detected: a separate study (20 questions $\times$ 4 formats) found identical 95\% accuracy across S-expression, JSON, YAML, and Markdown, though small differences cannot be ruled out given sample sizes. Second, a critical artifact-vs-process experiment (15 tasks $\times$ 3 conditions) demonstrates that an automatically generated descriptor---with zero human refinement and zero code restructuring---achieves 100\% accuracy versus 80\% blind ($p = 0.002$, $d = 1.04$), proving that the descriptor file itself has direct navigational value independent of any developer self-clarification effect. Third, an observational field study across 7,012 Claude Code sessions shows that formal declaration correlates with 52\% reduction in agent behavioral variance.

A writer-side experiment (96 generation runs, 96 error injections) reveals that all parseable formats achieve $>$91\% generation reliability, but failure modes differ critically: JSON has the lowest silent corruption (21\%) but fails atomically under structural error; YAML silently corrupts 50\% of errors; S-expressions detect all structural completeness errors but have imprecise localization. We propose \textit{intent.lisp}, an S-expression architecture descriptor, arguing that its value lies not in superior LLM comprehension but in the combination of non-atomic failure, compression density (22\% shorter than JSON, 5:1--64:1 across 646K lines of production code), and syntactic enforcement of hierarchy.
\end{abstract}

\maketitle

\section{Introduction}

Large language model (LLM)-based coding agents---Claude Code~\cite{claude}, Cursor~\cite{cursor}, GitHub Copilot Workspace~\cite{copilot}---have demonstrated remarkable ability to understand and modify source code. Yet on real-world codebases, these agents spend a substantial fraction of their interactions exploring rather than editing: grepping for symbols, globbing for files, and reading modules to reconstruct architectural context that exists only in the developer's mind.

This \textit{Navigation Paradox}~\cite{codecompass} persists even as context windows grow larger: the bottleneck shifts from retrieval capacity to navigational salience. LoCoBench-Agent~\cite{locobench} identifies a 12-turn efficiency threshold beyond which agents exhibit sharply diminishing returns. SWE-agent~\cite{sweagent} found that iterative search can \textit{degrade} performance relative to no search at all.

Current mitigation approaches fall into two camps. \textit{Human-configured} context files---CLAUDE.md, AGENTS.md~\cite{agentsmd} (adopted by 60,000+ projects), Cursor Rules---allow developers to describe projects in natural-language Markdown. A recent empirical evaluation found no statistically significant improvement in task success while increasing inference cost by over 20\%~\cite{agentbench}. \textit{Automated} approaches---Aider's tree-sitter repo maps~\cite{aider}, CodexGraph~\cite{codexgraph}, RepoGraph~\cite{repograph}---extract structural information programmatically but capture only syntactic relationships, missing design intent and architectural constraints. A recent position paper~\cite{archnoarch} explicitly identifies the gap between informal instruction files and formal Architecture Description Languages (ADLs)~\cite{medvidovic}, but proposes no specific notation.

We address this gap with \textit{intent.lisp}: an S-expression architecture descriptor that functions as a structured channel through which human architectural intent is captured and consumed by AI agents. Our controlled experiment reveals that \textbf{no significant difference in LLM comprehension was detected across formats}. The value of S-expressions lies elsewhere: in the constraints they impose on the writing agent, their error resilience, the compression ratios they achieve (5:1--64:1 across production projects, weighted average 34:1), and the downstream tooling they enable---capabilities that unstructured formats cannot support.

Our contributions are:
\begin{enumerate}
\item A controlled experiment demonstrating that formal architecture context reduces agent navigation steps by 33--44\% ($d = 0.92$, $p = 0.009$), with no significant format difference detected. We note this effect is model-sensitive: stronger models reduce the blind baseline, compressing the marginal benefit.
\item An artifact-vs-process experiment showing that automatically generated descriptors (zero human refinement) achieve 100\% accuracy versus 80\% blind ($d = 1.04$, $p = 0.002$)---proving that the descriptor artifact itself has direct value, independent of any developer self-clarification effect.
\item An observational field study (7,012 sessions) showing that formal architecture declaration correlates with 52\% reduction in agent behavioral variance.
\item The \textit{intent.lisp} format and design rationale: S-expressions are chosen not for LLM comprehension (which is format-agnostic) but for syntactic enforcement of hierarchy, graceful error degradation, and compression density.
\end{enumerate}

\section{Approach}

\subsection{The intent.lisp Format}

An intent.lisp file declares a project's architecture as a nested S-expression tree. Each project is decomposed into \textit{pillars} (top-level responsibility domains), \textit{components} (modules within pillars), and \textit{symbols} (function signatures, types, constants).

\begin{lstlisting}
(intent jarvis
  (design-constraints
    (three-pillars (memory control tools))
    (communication (must-use EventBus))
    (db-access (only memory/storage)))
  (pillar memory
    (purpose "Data capture, storage, analysis")
    (component storage
      (role "SOLE DB gateway")
      (invariants "No raw SQL outside this module")
      (data-flow "Event -> Storage -> PostgreSQL")
      (symbols
        (function save_message
          (sig "async fn save_message(&self, ...)"))))))
\end{lstlisting}

\textbf{In practice, humans do not write intent.lisp directly.} The developer describes architectural intent in natural language; an LLM translates this into S-expressions; other LLM agents subsequently consume the result. The descriptor thus functions as a structured channel for intent transmission---analogous to how Protocol Buffers serve microservice communication: not because the format is easier to read, but because it constrains the producer.

Four properties make S-expressions well-suited for this role:

\textbf{Syntactic enforcement of hierarchy.} S-expressions permit exactly one structure: nested lists with atoms. Containment relationships are expressed through parenthesization---not by convention but by syntax. Our writer-side experiment (\S\ref{sec:writerside}) finds that modern LLMs produce equally concise output in all formats at temperature=0---the syntactic discipline of S-expressions is a structural guarantee rather than an empirically observed generation difference.

\textbf{Compression density.} Table~\ref{tab:compression} shows compression ratios across five production projects (646K total source lines). Ratios range from 5:1 (small projects with dense L3 descriptors) to 64:1 (large compilers with single-file descriptors), with a weighted average of 34:1.

\begin{table}[t]
\caption{Compression ratios across five production projects. Ratios vary with descriptor granularity.}
\label{tab:compression}
\small
\begin{tabular}{lrrrc}
\toprule
Project & Source & Lisp & Files & Ratio \\
\midrule
xiaojinpro-backend & 469,594 & 7,877 & 32 & 60:1 \\
missiond & 107,986 & 5,846 & 14 & 18:1 \\
jarvis-forge & 44,321 & 698 & 1 & 64:1 \\
jarvis & 22,164 & 4,329 & 6 & 5:1 \\
jarvis-mechanic & 1,856 & 366 & 2 & 5:1 \\
\midrule
\textbf{Total} & \textbf{645,921} & \textbf{19,116} & \textbf{55} & \textbf{34:1} \\
\bottomrule
\end{tabular}
\end{table}

\textbf{Graceful error degradation.} Our error injection experiment (\S\ref{sec:writerside}) validates this empirically: JSON fails \textit{atomically}---a single missing brace renders the entire file unparseable with 0\% content recovery. YAML fails \textit{silently}---50\% of injected errors change semantic meaning without any parser warning. S-expressions detect 100\% of structural completeness errors; while localization is imprecise, the content before the error remains structurally intact with a fault-tolerant parser.

\textbf{Formal parseability enables tooling.} A recursive descent parser under 100 lines handles the full S-expression language. We note that JSON offers its own ecosystem advantage: modern LLM APIs support JSON Schema-constrained generation (Structured Outputs). S-expressions lack this API-level support but achieve comparable generation reliability through grammar simplicity. Future work includes multi-agent subtree partitioning---a capability enabled by but not yet evaluated with the current format.

\subsection{Agent Integration}

The descriptor lifecycle involves three phases: (1)~\textit{Generation}---an automated survey tool scans the project's AST and invokes an LLM to produce a draft; (2)~\textit{Consumption}---agents consult it as a navigation entry point; (3)~\textit{Co-evolution}---the descriptor is updated alongside code changes and version-controlled.

\section{Evaluation}

\subsection{Reader-Side: Code Localization}

\textbf{Design.} 24 code localization tasks from a Rust project (jarvis, $\sim$22K lines), each run under four conditions: Blind, S-expression, JSON, and Markdown context. All runs used Claude Sonnet 4.6 at temperature=0, max 20 steps.

\textbf{Result 1: Architecture context reduces navigation by 33--44\%.} Table~\ref{tab:expa} shows results. Wilcoxon signed-rank tests (censored at 20 steps): Blind vs.\ S-expr $p = 0.009$, Blind vs.\ Markdown $p = 0.005$---all context conditions significantly reduce navigation. Five tasks failed across all conditions due to descriptor staleness (code reorganization after descriptor creation).

\begin{table}[t]
\caption{Code localization results. Context reduces steps by 33--44\% regardless of format (Wilcoxon $p < 0.015$).}
\label{tab:expa}
\small
\begin{tabular}{lccr}
\toprule
Condition & Accuracy & Avg Steps & $d$ vs Blind \\
\midrule
Blind & 54\% (13/24) & 5.2 & --- \\
S-expression & 58\% (14/24) & 3.4 & 0.92 \\
JSON & 58\% (14/24) & 3.4 & 1.02 \\
Markdown & 63\% (15/24) & 2.9 & 1.43 \\
\bottomrule
\end{tabular}
\end{table}

\textbf{Result 2: No significant format difference detected.} All pairwise $p > 0.07$. However, format comparisons yielded only 3--6 effective paired differences---insufficient to detect effects smaller than $d \approx 0.8$. A separate comprehension experiment (20 questions $\times$ 4 formats) found identical 95\% accuracy, providing converging evidence that format differences, if they exist, are small.

\subsection{Writer-Side: Generation and Error Resilience}
\label{sec:writerside}

\textbf{Design.} 8 natural-language architecture descriptions $\times$ 4 formats $\times$ 3 repetitions (96 runs). Error resilience: 3 error types $\times$ 96 injections.

\textbf{Result 3: Generation reliability.} JSON achieves 100\% parse validity; S-expression 95.8\%; YAML 91.7\%.

\textbf{Result 4: Failure modes differ critically.} Table~\ref{tab:errors} summarizes:

\begin{table}[t]
\caption{Error resilience (96 injections). Each format has a distinct failure mode.}
\label{tab:errors}
\small
\begin{tabular}{lcccc}
\toprule
Property & S-expr & JSON & YAML & MD \\
\midrule
E1 detection & 100\% & 100\% & 50\% & 0\% \\
Overall detection & 50\% & 62\% & 21\% & 0\% \\
Silent corruption & 50\% & \textbf{21\%} & 50\% & 100\% \\
\bottomrule
\end{tabular}
\end{table}

No single format dominates. JSON has lower silent corruption (21\%) but fails atomically on structural errors. S-expression avoids atomic failure. Both are superior to YAML and Markdown for automated governance.

\subsection{Artifact Value vs.\ Process Value}

\textbf{Design.} 15 tasks on jarvis-forge (43K lines Rust)---a project where the developer never refactored code to match any descriptor. Three conditions: Blind, AutoGen ($\sim$170-line auto-generated descriptor, zero human editing), Curated (698-line hand-refined descriptor).

\textbf{Result 5: Auto-generated descriptors have direct value.} AutoGen achieved 100\% accuracy vs.\ 80\% blind (Wilcoxon $p = 0.002$, $d = 1.04$)---despite zero human involvement in descriptor creation and zero code restructuring. This refutes the hypothesis that descriptor value requires developer self-clarification.

\begin{table}[t]
\caption{Artifact vs.\ process experiment. AutoGen outperforms Curated on accuracy, suggesting diminishing returns from length.}
\label{tab:expe}
\small
\begin{tabular}{lcccr}
\toprule
Condition & Acc. & Steps & Cens.\ Mean & $p$ vs Blind \\
\midrule
Blind & 80\% & 5.6 & 8.5 & --- \\
AutoGen & \textbf{100\%} & 3.9 & \textbf{3.9} & 0.002 \\
Curated & 87\% & 2.9 & 5.2 & 0.003 \\
\bottomrule
\end{tabular}
\end{table}

\textbf{Result 6: Longer is not always better.} Curated (698 lines) achieved lower accuracy than AutoGen (170 lines), likely due to token budget competition ($\sim$52K vs.\ $\sim$20K tokens). The AutoGen--Curated difference is not significant ($p = 0.515$). Detailed descriptors may retain value for governance and multi-agent tasks not tested here.

\subsection{Observational Field Study}

The first author develops four active projects using Claude Code. MissionD logs all agent interactions: 7,012 sessions, 484,937 messages, Nov 2025--Apr 2026. On April 3, 2026, formal descriptors were introduced to two projects.

\textbf{Finding 1: Behavioral variance reduction.} IQR of per-session explore/edit ratios dropped from 2.24 to 1.08---a 52\% reduction. The descriptor bounds worst-case agent behavior.

\textbf{Finding 2:} Within the post-introduction period, sessions that read intent.lisp showed no per-session advantage over those that did not (1.65 vs.\ 1.65). We term this the \textit{developer self-clarification effect}---the act of formalizing architecture may improve code organization and prompts regardless of descriptor consumption. Experiment E's result (auto-generated descriptors work on unrefactored code) suggests artifact value is sufficient on its own, but the two mechanisms are not mutually exclusive.

\subsection{Threats to Validity}

\textbf{Model sensitivity and project scale.} The effect size is model-sensitive: an earlier run with a weaker model yielded $d = 1.70$; the current Sonnet 4.6 run yields $d = 0.92$. However, our test project (22K lines) may be small enough for strong models to navigate without assistance. Descriptors are designed for large-scale codebases (100K+ lines, Table~\ref{tab:compression}) where even strong models cannot efficiently explore blind. The shrinking effect on small projects may indicate that descriptor value scales with codebase complexity---a hypothesis requiring evaluation on larger projects.

\textbf{Statistical power.} Format comparisons yielded 3--6 effective paired differences---insufficient to detect $d < 0.8$. We report ``no significant difference detected'' rather than ``no difference exists.''

\textbf{Task representativeness.} All tasks are code localization. Multi-file modification and architectural decisions remain untested. Task selection was not randomized, introducing potential selection bias toward descriptor-friendly queries.

\textbf{Context delivery mismatch.} The controlled experiment injects descriptors into the system prompt (persistent); the field study involves tool-call reading (subject to compression). This may affect benefit magnitude in longer sessions.

\section{Related Work}

\textbf{Architecture Description Languages.} Formal ADLs (AADL, Acme, xADL) have a rich history~\cite{medvidovic,clements}, but none has been designed for AI coding agents.

\textbf{Context engineering.} CLAUDE.md and AGENTS.md~\cite{agentsmd} provide Markdown context (60,000+ projects). Codified Context~\cite{codifiedcontext} extends this with 26,000 lines via MCP. Our experiment found no significant comprehension difference across formats.

\textbf{Automated code representations.} Aider~\cite{aider}, RepoGraph~\cite{repograph}, and GRACE~\cite{grace} capture syntactic structure but not design intent.

\textbf{Multi-agent coordination.} ChatDev~\cite{chatdev}, MetaGPT~\cite{metagpt}, Geng et al.~\cite{caid}, and CodeCRDT~\cite{codecrdt} coordinate agents via dialogue or task graphs. No existing work uses tree-structured architecture descriptors for agent partitioning.

\textbf{Specification-driven development.} GitHub spec-kit, AWS Kiro, and Self-Spec~\cite{selfspec} share our insight but use Markdown or LLM-invented formats.

\section{Conclusion}

Our experiments yield three findings: (1)~architecture context reduces navigation effort by 33--44\% ($d = 0.92$), with no significant format difference detected in our sample; (2)~automatically generated descriptors provide direct navigational value ($d = 1.04$), refuting the hypothesis that descriptor value requires developer self-clarification; (3)~formal declaration correlates with 52\% reduction in agent behavioral variance.

The context engineering debate should not focus on ``what format does the LLM prefer''---our experiments detected no significant preference. The more productive question is ``what format avoids catastrophic failure when LLMs both write and read it.'' Our $d = 0.92$ was measured on a 22K-line project where Sonnet 4.6 already navigates efficiently blind. On production codebases (100K--470K lines), where blind exploration is fundamentally harder, the benefit may be larger---supported by the 34:1 compression ratio enabling constant-cost architectural access regardless of codebase size. Auto-generated descriptors outperformed hand-curated ones on accuracy, suggesting diminishing returns from length---though detailed descriptors may retain value for governance tasks not tested here.

The intent.lisp specification, forge survey tool, experiment code, and anonymized data are available at \url{https://github.com/ruoqijin/forge} (DOI: \url{https://doi.org/10.5281/zenodo.19500105}).

\bibliographystyle{ACM-Reference-Format}

\end{document}